\documentclass[aps,superscriptaddress,amssymb,showpacs,prl,twocolumn]{revtex4}
\usepackage{graphicx}
\begin{document}

\title{Quantum annealing of the Traveling Salesman Problem}

\author{Roman Marto\v{n}\'ak} 
\altaffiliation[Permanent address: ]{Department of Physics, 
  Slovak University of Technology, Ilkovi\v{c}ova 3, 812 19 Bratislava, 
  Slovakia}
\affiliation{Computational Science, Department of Chemistry and Applied
  Biosciences, ETH Z\"urich, USI Campus, Via Giuseppe Buffi 13, CH-6900
  Lugano, Switzerland}
\author{Giuseppe E. Santoro}
\affiliation{International School for Advanced Studies (SISSA) and 
INFM-Democritos, Trieste, Italy}  
\affiliation{International Center for Theoretical Physics (ICTP), P.O.Box 586,  Trieste, Italy}  
\author{Erio Tosatti}
\affiliation{International School for Advanced Studies (SISSA) and 
INFM-Democritos, Trieste, Italy}  
\affiliation{International Center for Theoretical Physics (ICTP), P.O.Box 586,  Trieste, Italy}  

\date{\today}

\begin{abstract}
  
  We propose a path-integral Monte Carlo quantum annealing scheme for the
  symmetric Traveling Salesman Problem, based on a highly constrained
  Ising-like representation, and we compare its performance against
  standard thermal Simulated Annealing.  The Monte Carlo moves implemented
  are standard, and consist in restructuring a tour by exchanging two links
  (2-opt moves). 
  The quantum annealing scheme, even with a drastically simple form of kinetic
  energy, appears definitely superior to the classical one, when tested on 
  a 1002 city instance of the standard TSPLIB.
\end{abstract}

\maketitle


Quantum annealing (QA) -- that is using quantum mechanics to optimize,
through annealing, hard problems of everyday life, including those
that have nothing to do with quantum mechanics -- is a relatively new and
fascinating idea, with important physical implications and potential impact
in a variety of areas, from technological applications to other disciplines
of science, wherever optimization of a complex system is the issue.

The idea of QA is an offspring of the celebrated thermal 
{\em simulated annealing} (SA) \cite{Kirk_CA,Cerny}, where the
problem of minimizing a certain cost (or energy) function in a large
space of configurations is tackled, through a statistical mechanics analogy,
by the introduction of an artificial temperature variable which is slowly 
reduced to zero in the course of a Monte Carlo (MC) or Molecular Dynamics 
simulation. 
This device allows to explore the configuration space avoiding trapping into 
local minima, often providing a more effective and less biased search for 
the minimal ``energy'' than standard gradient-based minimization methods.

Why not using quantum mechanics for the same purpose?
Quantum mechanics works with wave-functions that can equally well sample
wide regions of phase-space.
Instead of thermal fluctuations, one exploits here the 
{\em quantum fluctuations} provided by a suitably introduced 
-- and equally artificial -- kinetic energy.
Annealing is then performed by slowly reducing to zero the amount of quantum
fluctuations introduced.
Quantum fluctuations have, in many respects, an effect similar to that
of thermal fluctuations -- they cause, for instance, solid helium to melt
even at the lowest temperatures -- but they differ considerably in other 
respects. In particular, quantum systems can {\em tunnel through}
classically impenetrable potential barriers between energy valleys, 
a process that might prove more effective than waiting for them to be overcome
thermally, as in SA.

Formulated in the mid 90's \cite{Finnila}, the idea of QA
picked up momentum only recently, through experimental work such as that of 
Brooke {\it et al.} \cite{aeppli,aeppli2}, where it was shown that 
in the disordered Ising ferromagnet 
LiHo$_{0.44}$Y$_{0.56}$F$_4$, QA is both experimentally feasible and 
apparently superior to thermal annealing.
Stimulated by these findings we recently carried out a benchmark comparison
of classical and path-integral MC annealings on the two-dimensional random 
Ising model \cite{science,PRB_PIMC}.
In that study, we confirmed the superiority of QA. We also presented a 
simple theory based on Landau-Zener tunneling \cite{science} for that.
While other theoretical efforts had also reported some success with QA for
other problems \cite{QA_jpn,Berne1,Berne2,Farhi_science}, it is nonetheless
fair to say that our overall experience with tackling hard problems by QA
is still very limited.
The Traveling Salesman Problem (TSP), a classic hard optimization problem, 
provides an ideal playground for a further test of QA in comparison with SA.
In this Letter we report an application of QA to TSP, where we find that 
again it is superior to SA.

Given $N$ cities with set inter-city distances $d_{ij}$, 
TSP consists in finding the shortest route connecting them, visiting each city 
once and returning to the starting point. 
The literature on TSP is vast, and {\it e.g.} Ref.~\onlinecite{Johnson} 
can be consulted for an account of the algorithms proposed. 
SA, although never a winning scheme when compared to some of the ad-hoc 
algorithms specifically tailored to the TSP problem, 
is known to be a very flexible, simple, and competitive algorithm for the 
TSP as well \cite{Johnson}. 
The natural question is now: can QA do better than SA, with a comparable 
computational effort? We will now show that this is indeed the case.

The crucial point in devising a QA scheme is how to describe the Hilbert space 
of the problem, and how to write and implement 
a quantum Hamiltonian $H_{TSP}=H_{pot}+H_{kin}$. 
Here $H_{pot}$ represents the classical potential energy of a given 
configuration (in our case, the length of a tour), and $H_{kin}$ is a suitable 
kinetic energy operator providing the necessary quantum fluctuations, and 
eventually annealed to zero. 
The route we followed (certainly not unique, or even the most efficient) 
goes through the mapping to a highly constrained Ising-like system, somewhat 
reminiscent of the Hopfield-Tank mapping of TSP as a neural 
network \cite{Hopfield}. (See also Ref.~\onlinecite{Kadowaki}.)
Formally, each configuration of the system (a valid tour) is associated to a 
$N\times N$ matrix $\hat{T}$ with $0/1$ entries in the following way:
For every directed tour (an ordered sequence of cities) 
$\hat{T}_{i,j}=1$ if the tour visits city 
$i$ immediately after city $j$, and $\hat{T}_{i,j}=0$ otherwise. 
$\hat{T}$ has $N$ entries equal to $1$, all other elements 
being $0$, and obeys the typical constraints of a permutation matrix, i.e.:
{\it a)} All diagonal elements vanish, $\hat{T}_{i,i}=0$; 
{\it b)} There is a single $1$ in each row $i$ and in each column $j$. 
For the {\em symmetric} TSP problem we want to consider 
(a TSP where the distance matrix is symmetric $d_{ij}=d_{ji}$), 
a directed tour represented by a $\hat{T}$, and the 
reversed tour, represented by the transposed matrix $\hat{T}^t$, 
have exactly the same length. 
It is convenient, as will be apparent in a moment, to introduce the 
symmetric matrix $\hat{U}=\hat{T}+\hat{T}^t$ as representative of 
{\em undirected} tours, the non-zero elements of $\hat{U}$ being given by 
$\hat{U}_{i,j}=1$ if $i$ is connected to $j$ in the tour, $i$ being 
visited before or after $j$.
One can be readily convinced that there is no loss of information 
in working with $\hat{U}$ instead of $\hat{T}$. Given the matrix $\hat{U}$ 
there is no ambiguity in extracting the directed tour, 
represented by $\hat{T}$, which originated it.
The length of a tour can be expressed in terms of the $\hat{U}_{i,j}$, 
as follows:
\begin{equation}
H_{pot}(\hat{U}) = \frac{1}{2} \sum_{(ij)} d_{ij} \hat{U}_{i,j} 
= \sum_{\langle ij\rangle} d_{ij} \hat{U}_{i,j} \;,
\end{equation}
where $d_{ij}=d_{ji}$ is the distance between city $i$ and city $j$, and 
$\langle ij\rangle$ signifies counting each link only once. 
$H_{pot}$ is the potential energy we will finally seek to minimize. 

Since there is no natural physical kinetic energy in the problem we
have to devise a suitable one. This choice is arbitrary, and many
simple forms, such as transposition of two neighboring cities in a tour,
could provide one or another kind of quantum fluctuations.  Reasonably,
however, the choice of $H_{kin}$ should also encompass the important
elementary ``moves'' of the problem, determining which configurations are
to become direct neighbors of a given configuration, a choice which in turn
influences the problem's effective landscape \cite{theory_landascape}.
A very important move that is often used in heuristic TSP algorithms 
is the so-called {\em 2-opt move}, consisting in eliminating
two links in the current tour, $(c_1~\to~c_2)$ and $(c_{1'}~\to~c_{2'})$,
and rebuilding a new tour in which the connections are given by
$(c_1~\to~c_{1'})$ and $(c_2~\to~c_{2'})$. 
This is illustrated with a 8-city example in Fig.~1, were (left part)
the tour $(1\to 2\to 5\to 8\to 7\to 6\to 3\to 4)$, represented
by the matrices $\hat{T}_{\rm in}$ and $\hat{U}_{\rm in}$, 
is rebuilt by eliminating the two links 
$(c_1=2\to~c_2=5)$ and $(c_{1'}=3\to~c_{2'}=4)$, 
and forming (see right part of Fig.~1) two new links 
$(c_1=2\to~c_{1'}=3)$ and $(c_2=5\to~c_{2'}=4)$. 
The matrices $\hat{T}_{\rm fin}$ and $\hat{U}_{\rm fin}$ in the lower right part
of Fig.~1 represent the final (directed and undirected) tour obtained
after the 2-opt move. 
A 2-opt move implies that a whole section of the original tour 
(between the two removed links $(c_1\to c_2)$ and $(c_{1'}\to c_{2'})$ )
gets reversed in the new tour, yielding a long series of bit flips
in the matrix $\hat{T}$ (dotted circles in Fig.~1).
If we associate a spin variable $+1$ ($-1$) to each entry $1$ ($0$), 
and represent a 2-opt move in terms of spin-flip operators acting on the 
configuration $\hat{T}_{\rm in}$, we would need a whole string of spin-flip 
operators (a non-local object) to enforce a trivial operation 
-- reversing a piece of tour -- which does not affect the tour length at all.
The advantage of working with $\hat{U}$ is that it represents, in a symmetric 
way, the direct and the reverse tour, so that all the entries corresponding
to the section of reversed tour are completely untouched. 
The whole 2-opt move, when working with $\hat{U}$ matrices, can be represented 
by just four spin-flip operators:
\[
S^{+}_{\langle c_{1'},c_1\rangle} S^{+}_{\langle c_{2'},c_2\rangle} 
S^{-}_{\langle c_2,c_1\rangle} S^{-}_{\langle c_{2'},c_{1'}\rangle} \;,
\]
where, by definition, each $S^{\pm}_{\langle i,j\rangle}$ flips an Ising
spin variable (defined as $S^z_{\langle i,j\rangle}=(2\hat{U}_{i,j}-1)=\pm 1$)
at position $(i,j)$ and at the symmetric position $(j,i)$, i.e.,
$S^{\pm}_{\langle i,j\rangle}=S^{\pm}_{i,j} S^{\pm}_{j,i}$.
 
The quantum Hamiltonian for the TSP which implements the 2-opt moves is:
\begin{widetext}
\begin{equation} 
H_{TSP} = H_{pot}(\hat{U}) + H_{kin} 
= \sum_{\langle ij\rangle} d_{ij} 
              \frac{\left( S^z_{\langle i,j\rangle} + 1 \right)}{2} \;
- \frac{1}{2} \sum_{\langle ij\rangle} \sum_{\langle i'j'\rangle}
\Gamma(i,j,i',j';t) \; \left[ 
   S^{+}_{\langle i',i\rangle} S^{+}_{\langle j',j\rangle} \,
   S^{-}_{\langle j,i\rangle}  S^{-}_{\langle j',i'\rangle} 
+ H.c. \right] \;. \\
\label{H_TSP:eqn} 
\end{equation}
\end{widetext}
Here the quantum coupling $\Gamma$ is a real positive function depending, 
in principle, on the links, as well as, in the annealing problem, 
on time. 
The link dependence of $\Gamma$ can be restricted to a dependence on
the distances between the cities involved in the two links 
created by the 2-opt move, in such a way as to realize a 
{\em neighborhood pruning} \cite{Johnson}, by discouraging (or forbidding
altogether) the creation of links between distant cities.
%
\begin{figure}[!tbp]
\includegraphics*[width=7cm,angle=0]{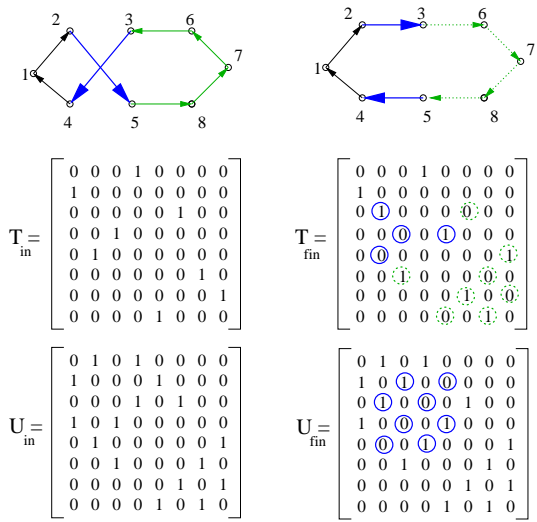}
\caption{ \label{fig:1} 
  Left: Representation of an 8-city tour, with the corresponding matrix
  $\hat{T}_{\rm in}$ and $\hat{U}_{\rm in}=\hat{T}_{\rm in}+\hat{T}_{\rm in}^t$.
  Right: The final tour obtained when a 2-opt move is performed, with
  a whole section reversed (dotted line).
  The matrices $\hat{T}_{\rm fin}$ and $\hat{U}_{\rm fin}$ 
  are shown, the circles indicating the entries that have been switched
  ($0\leftrightarrow 1$) by the 2-opt move. 
  The dotted circles in $\hat{T}_{\rm fin}$ are entries related to the trivial 
  reversal of a section of the tour.  
  }
\end{figure}
%
In so doing, we mapped the symmetric TSP problem onto a highly constrained
Ising spin problem with $N(N-1)/2$ sites -- one for each pair of $(i,j)$, 
with $i>j$, due to the symmetry $(i,j)\leftrightarrow (j,i)$,
which we denote by $\langle i,j\rangle$ --, with a four-spin-flip kinetic 
term providing the 2-opt quantum fluctuations.
The potential energy $H_{pot}$ appearing in $H_{TSP}$ (tour length),
represents, in the Ising language, a (random) external magnetic field at
each site $\langle i,j\rangle$ depending on the inter-city distance $d_{ij}$.  
The constraints on the Hilbert space are such that the matrix
$\hat{U}_{i,j} = (S^z_{\langle i,j\rangle}+1)/2$ represents a valid tour.  
In particular, the system lives in a subspace with a fixed magnetization 
-- exactly $N$ spins are up ($\hat{U}_{i,j}=1$) among the $N(N-1)/2$ -- 
and the 2-opt kinetic term conserves the magnetization.

As in the past \cite{science}, we will not attempt an actual Schr\"odinger 
annealing evolution of the quantum Hamiltonian proposed 
-- out of the question due to the large Hilbert space. 
On the contrary, we shall address the quantum problem by Quantum Monte Carlo 
(QMC), where annealing will take place in the fictitious ``time'' represented 
by the number of MC steps.  
Path-integral Monte Carlo (PIMC) provides a natural tool, due to its 
simplicity. However, attacking $H_{TSP}$ by PIMC meets with a first difficulty, 
namely Trotter discretization of the path-integral \cite{Suzuki}. 
That requires calculation of the matrix elements of the exponential operator
$\langle C'|\exp{-\epsilon H_{kin}}|C\rangle$ between arbitrary configurations 
$|C\rangle$ and $|C'\rangle$ of the system \cite{Suzuki}, 
a complicated affair for the $H_{kin}$ in Eq.~(\ref{H_TSP:eqn}). 
To circumvent this difficulty without giving up the simplicity of PIMC, 
we introduce a drastic simplification to our kinetic energy term, replacing 
it altogether with a standard transverse Ising form:
\begin{equation} 
\tilde{H}_{TSP} = 
    \sum_{\langle ij\rangle} d_{ij} 
      \frac{\left( S^z_{\langle i,j\rangle} + 1 \right)}{2} \;
- \Gamma(t) \sum_{\langle ij\rangle} [S^{+}_{\langle j,i\rangle} + H.c.] \;,
\label{simple_Hkin:eqn}
\end{equation} 
This form is trivially Trotter-discretized, as in the standard Ising system 
in transverse field \cite{Suzuki}, 
since the spin-flip term acts independently on single spins at each 
time-slice. 
Strictly speaking this simplified form of kinetic energy would no longer 
fulfill the constraint to take a valid tour to another valid tour. 
However, so long as we constrain the MC dynamics strictly within the valid 
tour subspace 
-- a restriction that comes automatically if we use exclusively 2-opt 
moves in the MC algorithm, and no single spin-flip moves -- 
that problem does not arise. 
That this way of bypassing the difficulty in treating the original Hamiltonian
(\ref{H_TSP:eqn}) should eventually produce a working QA scheme is a priori far
from obvious. We will eventually find that it indeed does.
The simplified single spin-flip kinetic term $\tilde{H}_{kin}$ 
in Eq.~(\ref{simple_Hkin:eqn}) 
enters only in calculating the weights of configurations of the Trotter 
discretized system, and not in the actual MC dynamics, which 
relies on the 2-opt moves and hence conserves the constraints. 
The details of the remaining implementation are identical to those reported 
for the random Ising case \cite{science,PRB_PIMC,pbc}. 
%
\begin{figure}[!tp]
\includegraphics*[width=8.5cm,angle=0]{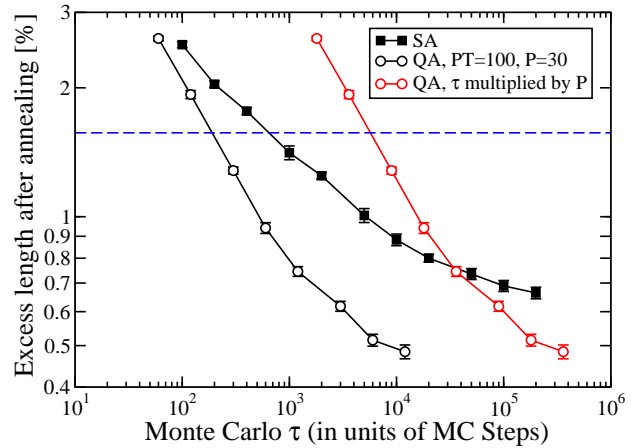}
\caption{\label{fig:2} 
  Average residual excess length found after Monte Carlo annealing for a
  total time $\tau$ (in MC steps), for the $N=1002$ instance pr1002 of the
  TSPLIB. Notice how quantum annealing (QA) provides residual excess lengths
  decaying faster than classical annealing (SA).
}
\end{figure}
 
For a direct test of our QA algorithm against SA we chose a standard benchmark
TSP problem, namely the printed circuit board instance 
pr1002 of the TSPLIB \cite{TSPLIB}. 
It is a structured TSP problem with $N=1002$ cities whose optimal tour length 
$L_{opt}$ is known exactly by other ad-hoc algorithms \cite{TSPLIB} to be 
$L_{opt}=259045$.
Our implementation of SA was a standard Metropolis MC with a temperature 
schedule starting from $T_0$ and going linearly to zero in a MC time $\tau$. 
We chose an optimal initial temperature $T_0$ by first performing several SA 
runs with different (short) cooling times $\tau$ starting from sufficiently 
high temperatures (the ergodic region).
This identifies an approximate ``dynamical temperature'' $T_{dyn}$ below 
which the cooling curves for different $\tau$'s start to differ.
For pr1002, we obtained $T_{dyn} \sim 100$. As it turns out, the optimal 
$T_0$ for SA approximately coincides with $T_{dyn}$, $T_0\sim T_{dyn}$, an 
implementation feature known for TSP SA as {\em cold starts} \cite{Johnson}.
For QA, we implemented for the Hamiltonian $\tilde{H}_{TSP}$ 
a similar PIMC to that used previously\cite{science,PRB_PIMC} 
at a fixed low temperature $T$ (we used $T=10/3$, see below). 
The quantum model is mapped onto a classical model with an 
extra imaginary-time dimension, consisting of $P$ ferromagnetically coupled 
replicas of the original spin problem, at temperature 
$PT$ \cite{Suzuki,PRB_PIMC} (we used $P=30$). 
Since QA requires initial configurations equilibrated at 
temperature $PT$ \cite{PRB_PIMC}, an obvious choice is to take $PT\sim T_{dyn}$,
i.e., $PT=100$ for the pr1002 \cite{pre-annealing}.
Finally, the transverse field $\Gamma$ is annealed linearly in a MC time $\tau$
from an initial value $\Gamma_0=300$ to a final value of zero. 
In both SA and QA, we used exclusively 2-opt moves, with a static
neighborhood pruning \cite{Johnson}, which restricts the attempted 2-opt
moves by allowing only a fixed number $M$ (we used $M=20$) of nearest
neighbors of city $j$ to be candidates for $j'$.
Our MC step consisted of $MN$ attempted 2-opt moves (for QA, in each of the
$P$ replicas).
In both SA and QA, we averaged the best tour length found over up to $96$ 
independent searches. 

Fig.~2 shows the results obtained for the residual error 
(average best-tour excess length) upon annealing for a total MC time $\tau$, 
$\epsilon_{exc}=(\bar{L}_{best}(\tau)-L_{opt})/L_{opt}$,
both with SA (filled squares) and with QA (open circles). 
As a reference, the best out of 1000 runs of the Lin-Kernighan algorithm 
(one of the standard local-search algorithms for TSP \cite{Johnson}) 
gives a percentage excess length of $\epsilon_{exc}^{LK} \approx 1.57 \%$ 
\cite{Johnson} (dashed line in Fig.~2).  
The results show that QA anneals more efficiently, reducing the error at a 
much steeper rate than SA.  
Moreover, even accounting for the extra factor $P$ in the total CPU time 
required by QA (rightmost open circles), QA is still more convenient than 
SA at large $\tau$ and small excess lengths. 
The similarity with the previous results on the random 2D Ising magnet
\cite{science} is striking. 
We argued in Ref.\ \cite{science} that QA is faster for the random Ising
case due to the superior ability of quantum physics to cross barriers through 
Landau-Zener tunneling, as compared to classical physics requiring for them 
to be overcome thermally: Such feature, although by no means a general
property, is apparently shared by the TSP.  
The upward curvature of the data in Fig.\ 2 is likely to signal a 
logarithmically slow annealing for both SA and QA \cite{science}. 
Also worth mentioning is the effect of the finite value of $P$, which is 
likely to be responsible for a saturation effect of the QA data, as shown in 
Fig.\ 1 of Ref.\ \onlinecite{science} for the random Ising case.

Current ad-hoc TSP algorithms do better than either annealings 
-- SA and QA being generic tools -- for the same CPU time.
Moreover, it is known that combining SA with local search heuristics
provides superior results to pure SA for the TSP problem \cite{Martin}.
Nevertheless, the absolute quality of QA and its success in a fair
comparison to SA strongly encourages to further applications of QA as a
general purpose optimization technique, and to possible improvements of the
bare QA scheme by combining it with other local heuristics, in the spirit
of Ref.\ \cite{Martin}.
Equally instructive will be to experiment with other QA schemes, for instance 
Green's Function QMC, which are able to cope with the 2-opt $H_{kin}$ 
constructed in Eq.~(\ref{H_TSP:eqn}) or with other sources of quantum 
fluctuations. That broadening of the project we must leave for future studies.
We believe that gaining further experience with the effects of artificially 
introduced quantum fluctuations in classical complex problems represents 
a very promising and challenging route, particularly in view of future 
developments in quantum computation.

\begin{acknowledgments} 
This project was sponsored by MIUR, through COFIN and FIRB, and by
INFM/Supercalcolo grant. 
R.M. acknowledges the hospitality provided by SISSA and ICTP. 
We thank D. Battaglia, C. Micheletti, and R. Zecchina for discussions.
\end{acknowledgments}



\begin{thebibliography}{99}
\bibitem{pbc} Here we also relax the periodic boundary condition along the
  imaginary-time direction. 
\bibitem{pre-annealing}
We initialize both the SA and the QA Trotter
configurations by performing short SA pre-annealings from a temperature
of $5T_0$ down to $T_0$, see Ref.~\protect\onlinecite{PRB_PIMC}.
\end{thebibliography}
\end{document}